\documentclass[aip,reprint]{revtex4-1}
\draft 
\usepackage{graphicx}
\usepackage{color}

\begin{document}


\title{Highly sensitive detection of magneto-optical markers based on magneto-optical gate effect} 

\author{Xinchao Cui}
\affiliation{School of Artificial Intelligence and Automation, Huazhong University of Science and Technology, Wuhan 430074, China.}

\author{Chong Lu}
\affiliation{Tongji Medical College, Huazhong University of Science and Technology, Wuhan 430022, China.}

\author{Chunping Liu}
\affiliation{Tongji Medical College, Huazhong University of Science and Technology, Wuhan 430022, China.}

\author{Wenzhong Liu}
\email[]{lwz7410@hust.edu.cn}
\altaffiliation{He is also with the China-Belt and Road Joint Laboratory on Measurement and Control Technology, Huazhong University of Science and Technology, Wuhan 430074, China and Shenzhen Huazhong University of Science and Technology Research Institute, Shenzhen 518063,China.}
\affiliation{School of Artificial Intelligence and 	Automation, Huazhong University of Science and Technology, Wuhan 430074, China.}

\date{\today}

\begin{abstract}
In this paper, a new concept of magnetic nanoparticles as magneto-optical markers is proposed, and a “magneto-optical gate” effect is explored. High-sensitivity detection of magnetic nanoparticles (MNPs) was developed based on this principle. Under irradiation from monochromatic light with a wavelength much larger than the size of a given MNP, Rayleigh scattering occurs on the MNP surface. The MNPs anisotropic and orientate themselves into chains, meaning that the Rayleigh scattering now decreases under magnetic field excitation. After applying an AC magnetic field of frequency $ f $, the transmitted light passing through the MNPs reagent generates an AC signal with a frequency of 2$ f $. Moreover, the 2$ f $ signal disappears after removal of the magnetic field. This creates a “magneto-optical gate” effect. The instantaneous and highly sensitive detection of magneto-optical markers with a concentration of 0.2 $ \mu $g/mL  and an effective optical path length of 10 $ \mu $m is realized based on this “magneto-optical gate” effect.
\end{abstract}

\pacs{}
\maketitle 

\section{Introduction}
Magnetic nanoparticles (MNPs) have been used as magnetic contrast agents in medical imaging for many years\cite{RN559,RN560} and can also obtain temperature information of targets through their temperature sensitivity\cite{RN19}. Under a static magnetic field, MNPs change the local magnetic field, which affects the contrast in MRI images\cite{RN561,RN567}. Gleich \textit{et al}. proposed magnetic particle imaging (MPI) based on the special performance of MNPs superparamagnetism under AC magnetic field excitation\cite{RN233}. MPI, as a functional imaging method, could have great potential in medical diagnosis\cite{RN469}. Importantly, MNPs are currently involved in various fields of biomedicine, including cell tracking, magnetic immunoassay, drug delivery, and tumor hyperthermia\cite{RN565,RN566,RN304,RN268,RN568,RN577,RN571,RN578}.

However, the current mechanisms involved magnetic detection and magnetic localization hinder further improvements in the sensitivity and resolution of MNP detection devices. Water and organic molecules gradually dominate the reagents and show diamagnetism with decreasing MNP concentration both in vitro or in vivo\cite{RN547}. The diamagnetism affects the magnetic response signal of MNPs, especially for fundamental harmonics\cite{RN546}. The high-order harmonics of MNPs can effectively avoid this diamagnetism under AC magnetic field excitation\cite{RN142}, but they are still indistinguishable from the ferromagnetic background of the device\cite{RN237}. Therefore, the detection sensitivity of MNPs is easily limited by the electromagnetic induction detection principle. In addition, the local magnetic field generated by MNPs reduces the localization accuracy under a magnetic field gradient\cite{RN575}. For existing MRI and MPI devices, the magnetic localization accuracy is stagnant at the scale of 500 $ \mu $m, and it is difficult to break through 100 $ \mu $m. In other words, the current spatial resolution and detection sensitivity of magnetic imaging make it difficult to detect cells in vivo\cite{RN580}.

MNPs as magneto-optical markers is a recent concept and is expected to break the ceiling of magnetic imaging resolution. This would extend the use of MNPs to high-resolution cellular-scale imaging. MNPs can be used as magneto-optical markers thanks to their special optical properties. Rayleigh scattering occurs on the MNP surface under irradiation from light with a wavelength much larger than the particle’s size. The scattering phenomenon of randomly distributed nanoparticles has an attenuating effect on the transmission of light passing through the MNPs reagent. In particular, MNPs align to form chains under magnetic field excitation, significantly reducing the Rayleigh scattering. The optical signal transmitted passing through the MNPs reagent will carry information on their size, concentration, and even temperature under magnetic field modulation. Moreover, higher resolution and increased sensitivity  detection of MNPs can be achieved by optical signal.

In this paper, we discuss a novel “magneto-optical gate” effect exclusively for the detection of magneto-optical markers. Magneto-optical marker detection is a method combines magnetic field modulation and optical detection using the magnetic and optical effects of MNPs. Monochromatic light is used as a carrier signal to irradiate the sample containing magneto-optical markers, and an alternating magnetic field is applied in parallel with the light. Our experimental results show that, after applying an AC magnetic field with frequency $ f $, light passing through the MNPs reagent generates an AC signal with frequency 2$ f $ . Moreover, the 2$ f $ signal disappears after removal of the magnetic field. This indicates that the AC magnetic field has a direct causal relationship with the second harmonic of the optical signal; that is, the magneto-optical markers have a “magneto-optical gate” effect. This “magneto-optical gate” effect is studied experimentally, focusing on MNP particle size, concentration, and optical path length. Finally, a highly sensitive, nanometer-scale, magneto-optical marker detection method based on active optical excitation is realized.

\section{Methods and materials}
\subsection{Principle}
MNPs exhibit anisotropy under the action of an external magnetic field and aggregate and align along the direction of the field to form chains with a cross-sectional diameter of about a few $ \mu $m. In this scenario, the intensity of light transmitted through the magnetic fluid changes with application of an external magnetic field\cite{RN480}. The scattering effect of particles under a magnetic field is greatly reduced compared with the scattering effect of particles in a random distribution state; therefore, the transmittance of the light passing through the MNPs reagent increases. This phenomenon can be explained by using a Monte Carlo model, and the simulation results are in good agreement with the experimental data\cite{RN483}. In addition, experiments show that MNP transmittance has a linear relationship with MNP concentration and MNP surface modification\cite{RN493}.

\begin{figure}[htbp]
	\centering
	\includegraphics[width=0.9\linewidth]{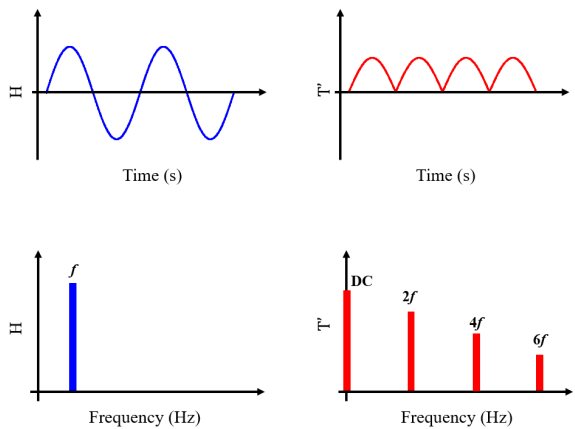}%
	\caption{\label{fig:figure1} Schematic diagram of MNP detection principle based on magneto-optical gate effect.}
\end{figure}

In this research, we detect changes in the transmittance by using AC magnetic field excitation to suppress background noise. The light passing through the MNPs reagent is modulated to a known frequency and the signal-to-noise ratio (SNR) is improved by phase sensitive detection (PSD). Moreover, a modulated optical signal avoids issues concerning the stability of the laser and the detection device. Notably, the transmittance of pure water is only related to the wavelength of the transmitted light and does not change with changes in the magnetic field. Furthermore, changes in transmittance due to Rayleigh scattering are independent of magnetic field direction, i.e., it is reciprocal. This means that changes in the intensity of light transmitted through MNPs reagent appear as even harmonics, e.g. DC, 2$ f $, 4$ f $ and 6$ f $, under a sinusoidal magnetic field excitation with frequency $ f $.

If $ I_{0} $ is the incident light intensity and $ I_{out} $  is the outgoing light intensity, the specific transmittance of light passing through MNPs reagent is $ T'= I_{0}/I_{out} - 1 $ under a sinusoidal magnetic field excitation $ H = H_{0} \sin(\omega t) $, as shown in Fig. \ref{fig:figure1}. $ T'  $ depends only on the magnitude of the magnetic field, not on its direction. Therefore, under modulation by sinusoidal magnetic field excitation $ H $, $ I_{out} $ is given by 
\begin{eqnarray}\label{equ:euqation1}
	I_{out}   &=&I_{0} \left(1+T' \left|\sin(\omega t)\right|\right) \nonumber\\
		      &=&I_{0} \left(1+T' \left(A_{0}+\sum_{i=1}^{2i} A_{2i} \sin(2i\omega t)\right)\right),
\end{eqnarray}
where $ A_{0} $ is the DC component coefficient of the Fourier expansion of $ T' $ and $ A_{2i} $ are the coefficients of even harmonics. The influence of a paramagnetic or diamagnetic background can be avoided, in essence, by measuring these harmonics.

\subsection{Experimental setup}
\begin{figure}[htbp]
	\centering
	\includegraphics[width=0.9\linewidth]{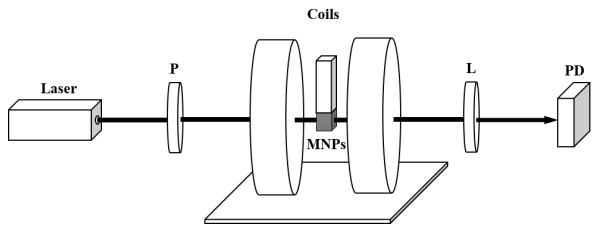}%
	\caption{\label{fig:figure2} Schematic diagram of MNP detection device based on magneto-optical gate effect. A laser produces 520 nm linearly polarized light I0; P is a polarizer used to adjust the light intensity; L is a plano-convex mirror used to focus the laser beam to a spot; PD is a photodetector.}
\end{figure}

A schematic diagram of the magneto-optical marker detection device is shown in Fig. \ref{fig:figure2} The sinusoidal modulation signal is generated by a data acquisition card (USB6356, NI), and then amplified by a linear power amplifier (7224, AE Techron) to drive a Helmholtz coil (FE-1FM180, Hunan Forever Elegance Technology). This generates an excitation magnetic field. Helmholtz coils have a flexible operating space and provide a sufficiently uniform magnetic field (the uniformity in a cube with a side length of 23 mm can reach 0.1\%). A continuous 520 nm laser (OM-12A520-10-G, Oeabt) was used as a light source, and the light illuminate and penetrate the sample parallel to the magnetic field direction. The laser was polarized, so that $ I_{0} $ could be adjusted by adding a polarizer. A lens was used to reduce the spot diameter to improve detection accuracy. $ I_{out} $ was converted to a voltage signal by a low noise photodetector (PDA8A2, Thorlabs) and read out using a data acquisition card (USB6356, NI).

The even harmonic decreases with the increase of order, so the second harmonic with the largest SNR was selected as the detection signal. The total noise of the device comprised mainly the 1/f noise of the magnetic field generation device, the Poisson noise caused by the random fluctuation of photons, and the inherent Johnson noise of the components. The Helmholtz coil had a large inductive reactance due to its large physical radius (105.64 mm), which limited the magnetic field excitation frequency. The magnetic field modulation frequency was finally chosen to be 90 Hz to make the signal as unaffected by 1/f noise as possible. The Johnson noise could not be eliminated, but it was much less than the Poisson noise and therefore could be ignored. The Poisson noise was also unavoidable, but increasing the light intensity as much as possible within the dynamic range allowed by the data acquisition card can improve the SNR. The data recording time of the following experiments is 1 s.

\subsection{Nanoparticles and container}
 The magneto-optical markers used for the experiment were water-based iron oxide nanoparticles (Ocean NanoTech) modified on the surface with carboxyl, which have excellent colloidal stability. Specific information on the size of the MNPs is shown in Table \ref{tab:table1}. The hydrodynamic particle size was measured by a particle size analyzer (Zetasizer Nano ZEN3690, Malvern) using the dynamic light scattering (DLS) method. Magneto-optical markers were placed in a quartz cuvette to ensure good transmission in the wavelength range: 200-2500 nm.

\begin{table}[htbp]
	\caption{\label{tab:table1}Core size and hydrodynamic diameter of different particles.}
	\begin{ruledtabular}
		\begin{tabular}{ccc}
			MNPs Model &  Core size(nm) & Hydrodynamic diameter(nm) \\ 
			\hline
			SHP-10 & 10 & 24.36 \\
			SHP-15 & 15 & 24.36 \\
			SHP-20 & 20 & 43.82 \\
			SHP-25 & 25 & 28.21 \\
			SHP-30 & 30 & 32.67 \\
			MHP-50 & 50 & 58.77 \\
		\end{tabular}
	\end{ruledtabular}
\end{table}

\section{Results and discussion}
\begin{figure}[htbp]
	\centering
	\includegraphics[width=0.9\linewidth]{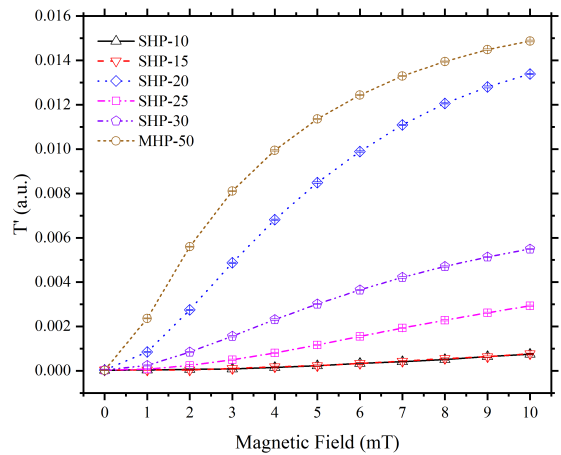}%
	\caption{\label{fig:figure3} Measurement results of $ T' $ vs. MNPs size. Symbols are experimental data, lines are guides to the eye, whereas error bars represent standard deviations.}
\end{figure}

Magneto-optical markers with different particle sizes were selected for experiments to verify the effect of particle size on detection. Magneto-optical markers were diluted to 100 $ \mu $g/mL (iron concentration) to ensure good light transmittance. Fig. \ref{fig:figure3} shows that $ T' $ of the MNPs increases with increasing external magnetic field, but it tends gradually plateau with further increase of the magnetic field. This is due to the aggregation of MNPs under the action of the magnetic field. The number of freely dispersed MNPs decreases, as does the degree of light scattering; therefore, the intensity of transmitted light gradually increases. However, the magnetic nanoparticles aggregate to their maximum extent, maintaining a stable state, with further increase of the magnetic field, i.e., $ T' $ tends to saturate.

The larger the hydrodynamic particle size of a MNP for a given concentration is, the larger $ T' $ is under the action of the magnetic field. In particular, SHP-20 has the largest $ T' $ apart from MHP-50 due to its large hydrodynamic particle size. Moreover, as the hydrodynamic particle size of a MNP increases, $ T' $ begins to saturate. When the wavelength of the incident light is much larger than the particle size, Rayleigh scattering occurs under irradiation of the surfaces of uniformly dispersed magneto-optical markers. The scattered light intensity is proportional to the number of particles and the 6th power of the particle diameter. Therefore, the larger the hydrodynamic particle size, the larger the magnetic nanoparticle $ T' $. However, particles with a larger particle size per unit volume are fewer for a given concentration. As a result, they are more likely to reach a stable state under the action of a magnetic field, and $ T' $ is more likely to saturate.

\begin{figure}[htbp]
	\centering
	\includegraphics[width=0.9\linewidth]{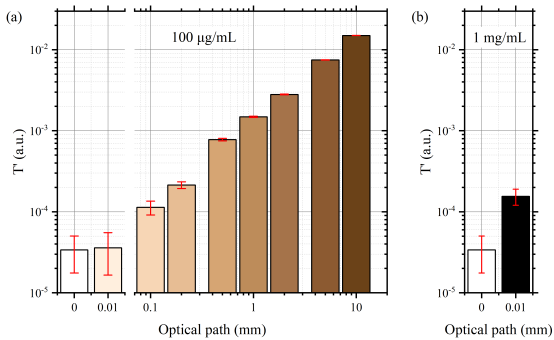}%
	\caption{\label{fig:figure4} Measurement results of $ T' $ vs. optical path. (a) Experimental results of 100 $ \mu $g/mL MNPs. (b) Experimental results of 1 mg/mL MNPs. The optical path 0 is the control group of the empty signal.Columns are experimental data, whereas error bars represent standard deviations.}
\end{figure}

The sample MHP-50, which produced the largest signal, was prepared at a concentration of 100 $ \mu $g/mL and placed in cuvettes with different thicknesses, from 10 mm to 10 $ \mu $m, to further explore the relationship between $ T' $ and the optical path, under a magnetic field excitation of 10 mT. Fig. \ref{fig:figure4}(a) shows that $ T' $ decreases proportionally with cuvette thickness. Limited by the detection noise of the whole system, the effective optical path that 100 $ \mu $g/mL MNPs can detect is 100 $ \mu $m. Using a higher concentration of 1 mg/mL MNPs allows the detection of an effective optical path of 10 $ \mu $m, (Fig. \ref{fig:figure4}(b)). This method has no optical path length limitation in theory. As long as MNPs are present, the effect on light scattering under the action of a magnetic field can be reduced, resulting in a corresponding change in transmitted light intensity. 

\begin{figure}[htbp]
	\centering
	\includegraphics[width=0.9\linewidth]{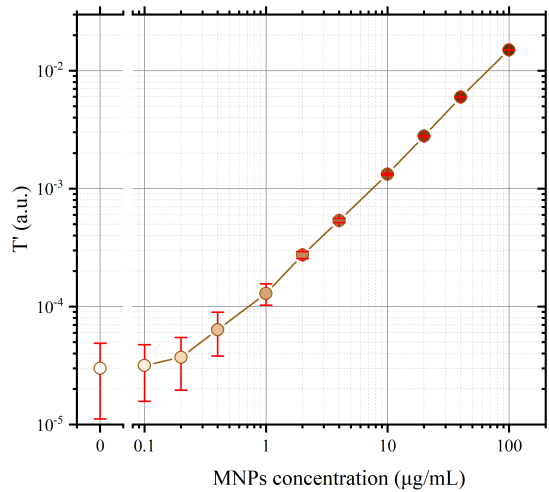}%
	\caption{\label{fig:figure5} Measurement results of $ T' $ vs. MNPs concentration. Concentration 0 is the control group of pure water.Symbols are experimental data, lines are guides to the eye, whereas error bars represent standard deviations.}
\end{figure}

The lower limit of detection for MNPs determines the depth and breadth of their application in the biomedical field. For MNPs to be extended to in vivo clinical use, their iron concentration must be at least below the safety standard of about 40 $ \mu $mol/L (about 2$ \mu $g/mL)\cite{RN293}. Previous studies have shown that MNP signals can be severely affected by the diamagnetism of water at low concentrations\cite{RN547,RN546}. Therefore, we examined the effect of MNP concentration on $ T' $ by diluting the MHP-50 sample from 100 $ \mu $g/mL to 0.1 $ \mu $g/mL. A standard quartz cuvette with an optical path of 10 mm was used under a magnetic field excitation of 10 mT, and a pure water control group was used. Fig. \ref{fig:figure5} shows that $ T' $ is only related to (is strictly proportional to) MNP concentration. Therefore, $ T' $ can be used as a detection index to quantitatively analyze the concentration of MNPs in a high diamagnetic background. The lower limit of detection for MNPs was 0.2 $ \mu $g/mL within the error range of the systematic measurements.

The “magneto-optical gate" detection method originates from the Rayleigh scattering of light by magneto-optical markers, so the intensity of $ T’ $ depends on the number of particles. The experimental results also show that $ T' $ is proportional to the optical path and the concentration of MNPs, that is, proportional to the number of particles. In particular, the method does not exhibit nonlinearity for short optical paths and low concentrations, which is conducive to the clinical application of MNPs. Therefore, this method holds promise for MNP detection in cells or tissue sections, and has great potential for the detection of trace amounts of MNPs.

\section{Conclusion}
This paper proposes the concept of magnetic nanoparticles as magneto-optical markers and discusses the “magneto-optical gate” effect using magneto-optical markers. MHP-50 were selected as magneto-optical markers to perform experiments with different optical paths and different concentrations. The amount of magneto-optical markers was proportional to the second harmonic signal of $ T' $ under a magnetic field excitation frequency $ f $. The “magneto-optical gate” effect realized a quantitative trace detection of 0.2 $ \mu $g/mL MNPs. The gate effect also allowed the accurate detection of the MNPs reagent with a thickness of 10 $ \mu $m on the self-built equipment. The detection of MNPs concentration and thickness has the potential to be further improved. In future work, magneto-optical markers are expected to be used for the detection of trace viruses or for functional imaging in vivo.


%
%

%

\begin{acknowledgments}
This work was supported by the MOST (Grant No. 2022YFE0107500), the Key Project of Hubei Province (Grant No. 2021ACB001), the Science, Technology  and Innovation Commission of Shenzhen Municipality (Grant No. JCYJ20210324142004012 and JCYJ20200109110612375).
\end{acknowledgments}

\bibliography{your-bib-file}

\end{document}